# Giant Acoustic Geometric Spin and Orbital Hall Effect

Wei Wang, Yang Tan, Jingjing Liu[†], Bin Liang[†], Jianchun Cheng[†]

Key Laboratory of Modern Acoustics, MOE, Institute of Acoustics, Department of Physics, Collaborative Innovation Center of Advanced Microstructures, Nanjing University, Nanjing 210093, People's Republic of China

[†]Correspondence and requests for materials should be addressed to Jingjing Liu (email: liujingjing@nju.edu.cn), Bin Liang (email: liangbin@nju.edu.cn), or Jianchun Cheng (email: jccheng@nju.edu.cn)

**Abstract**

Acoustic waves in fluid with spin-0 nature have been long believed not to support spin Hall effect and strong orbital Hall effect that enables experimental observation. Here we report the first theoretical explication and experimental demonstration of giant acoustic geometric spin and orbital Hall effect characterized by a large transverse shift. We reveal that this effect occurs when a vortex beam is observed from a tilted reference frame free of wave-interface interactions or gradient-index media needed for observing conventional ones, and can be amplified by simply binding the beam tightly. Thanks to this mechanism, large transverse shifts proportional to angular momentum are observed in a compact system. Our work provides deeper insights into the physics of angular momentum of classic waves.

Acoustic vortex beam, as a typical longitudinal scalar wave beam carrying orbital angular momentum, has been studied in depth both in theory and applications [1-7]. Recent findings demonstrate that the acoustic orbital Hall effect, as an acoustical orbital counterpart of conventional photonic spin Hall effect, can occur during reflection/refraction of a paraxial vortex beam by a metasurface or a sharp interface between two media, or while propagating in a gradient-index medium [8,9]. Nonetheless, the transverse shift is typically only a fractional order of magnitude of the wavelength, in contrast to the diameter of the acoustic beam which needs to be substantially larger than the wavelength to satisfy the paraxial approximation [8,9]. This implies that detecting this phenomenon is tremendously challenging without precise observational methods of quantum weak measurement [10] and the mechanism for effectively amplifying the transverse shifts [11] which play significant roles in observing the photonic Hall effect.

On the other hand, while nonparaxial acoustic beams have been recently shown to carry non-zero local spin density even in fluids, the integral spin angular momentum of localized acoustic wave in free space vanishes in agreement with the spin-0 nature of the longitudinal phonons [12-16]. Hence, those spin Hall effects in optics, which occur in paraxial circularly polarized light beams in various scenarios including propagation through gradient-index media [17], refraction/reflection at interfaces [18], or changes of observation coordinate systems [19], do not have direct acoustical spin counterparts. Consequently, despite the extensive research conducted on the photonic spin Hall effect and its vast potential in diverse applications [10,11,17-22], the fundamental theory of acoustic spin Hall effect in fluids remains an uncharted territory.

In this Letter, we give the first theoretical explication and experimental observation of the acoustic geometric spin and orbital Hall effect in fluid, which does not require the sound-interface interactions or the gradient-index media as conventional ones, but results from the equivalent spin-orbit and orbit-orbit interactions induced by the change of the observation coordinate system. This effect occurs when a nonparaxial acoustic vortex beam is observed from a tilted reference frame relative to the propagation axis, manifested by the transverse shift of the normal linear momentum density centroid, as shown in Fig. 1. The magnitude of spin- and orbital-induced transverse shifts exhibit a direct proportionality with the corresponding spin and orbital angular momenta, respectively, resulting in the total transverse shift proportional to the total angular momentum. Remarkably, we reveal that such an effect can be amplified by simply binding the beam transversely, which at the same time eliminates the interference factor induced by beams' amplitude/phase variation. Based on this mechanism, we design a compact platform supporting diffractionless vortex modes and accurately observe giant acoustic Hall effect in experiment with large transverse shifts by beams whose diameter are smaller than the wavelength without any additional interfaces that would otherwise arouse extra systematic errors , which overcomes the dependence on an auxiliary polarizer and the sensitivity to system errors in the observation of optical geometric spin Hall effect [23].

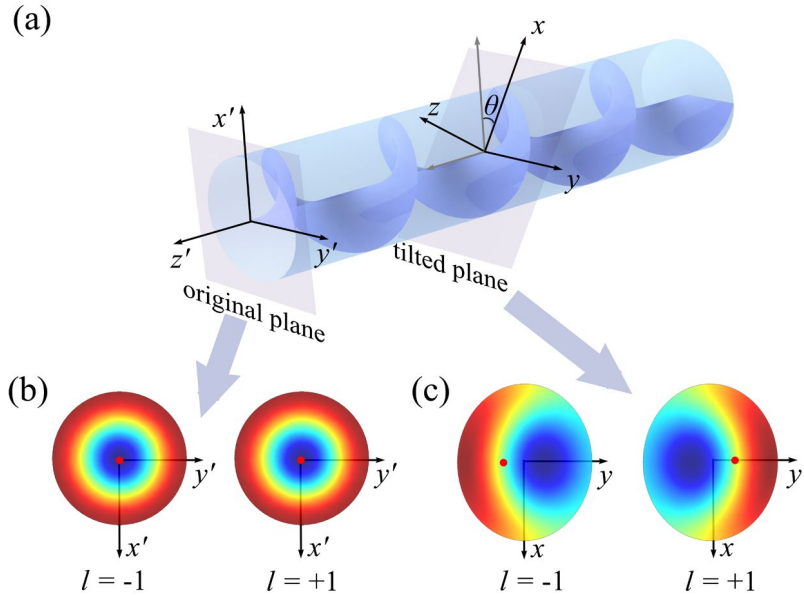

FIG. 1. (a) Schematic of acoustic geometric spin and orbital Hall effect. Normal linear momentum density distributions at the (b) original plane and (c) tilted plane, where red dot is the location of the centroid.

Consider an acoustic vortex beam propagating in cylindrical coordinate system

$(\xi,\varphi,z)$ along $z$-axis, its typical acoustic pressure can be written as $p(\xi,\varphi,z;t) = A(\xi,z)\exp\left[i\left(l\varphi + k_z z - \omega t\right)\right]$, where $A(\xi,z)$ is the complex amplitude of the beam, $\omega$ is the circular frequency, $k_z$ is the axial wavenumber satisfying $k_\xi^2 + k_z^2 = k^2$ with $k_\xi$ and $k = 2\pi/\lambda$ being the radial wavenumber and the wavenumber respectively. Given that diffraction phenomena result in the extreme difficulty of observing spin/orbital Hall effect in free space, the ideal beam used to observe this effect should be diffractionless with its envelope being unchanged during propagation. This means that $A(\xi,z) = A_{|l|}(\xi)$ and $A_{|l|}(\xi)$ is a real function describing the radial amplitude distribution of the beam with $l = 0, \pm 1,\ldots$ being the azimuthal quantum number. Hence, the acoustic pressure of the beam can further be expressed as

$$p(\xi,\varphi,z;t) = A_{|l|}(\xi)\exp\left[i\left(l\varphi + k_z z - \omega t\right)\right]. \tag{1}$$

Then the cycle averaged total linear momentum density (kinetic) in Cartesian coordinate can be obtained as follows

$$\mathbf{g}^t(\mathbf{r}) = \frac{1}{T}\int_0^T \rho(\mathbf{r};t)\mathbf{v}(\mathbf{r};t)dt = \frac{1}{2c_0^2\rho_0\omega}\left[A_{|l|}\left(\sqrt{x^2+y^2}\right)\right]^2\left(\frac{-ly}{x^2+y^2}\hat{\mathbf{x}} + \frac{lx}{x^2+y^2}\hat{\mathbf{y}} + k_z\hat{\mathbf{z}}\right),\tag{2}$$

where $\mathbf{r} = x\hat{\mathbf{x}} + y\hat{\mathbf{y}} + z\hat{\mathbf{z}}$ is the position vector, $c_0$ and $\rho_0$ are the sound speed and density of the media, respectively; $\rho(\mathbf{r};t)$ and $\mathbf{v}(\mathbf{r};t)$ are the mass density and velocity field, respectively. Kinetic momentum density can be divided into two parts: canonical momentum density related to orbital angular momentum and the rest part related to the spin angular momentum [12], expressed as

$$\mathbf{g}^o(\mathbf{r}) = \frac{1}{4\omega}\mathrm{Im}\left[\beta p^*\nabla p + \rho_0\mathbf{v}^*\cdot(\nabla)\mathbf{v}\right] = \frac{1}{2\rho_0\omega}\left(\frac{B_{|l|}(\zeta)}{2\omega^2} + \frac{1}{c_0^2}\left[A_{|l|}(\zeta)\right]^2\right)\left(-\frac{ly}{\zeta^2}\hat{\mathbf{x}} + \frac{lx}{\zeta^2}\hat{\mathbf{y}} + k_z\hat{\mathbf{z}}\right),$$

$$\mathbf{g}^s(\mathbf{r}) = \frac{\rho}{4\omega}\nabla\times\mathrm{Im}\left(\mathbf{v}^*\times\mathbf{v}\right) = -\frac{B_{|l|}(\zeta)}{4\rho_0\omega^3}\left(-\frac{ly}{\zeta^2}\hat{\mathbf{x}} + \frac{lx}{\zeta^2}\hat{\mathbf{y}} + k_z\hat{\mathbf{z}}\right),$$

$$\tag{3}$$

where $B_{|l|} = \left[A'_{|l|}(\zeta)\right]^2 + A_{|l|}(\zeta)A''_{|l|}(\zeta) - A_{|l|}(\zeta)A'_{|l|}(\zeta)/\zeta$, $\zeta = \sqrt{x^2 + y^2}$, and $A'_{|l|}(\zeta) = dA_{|l|}(\zeta)/d\zeta$. The centroid shift of these three kinds of linear momentum density on the observation plane can be expressed as follows

$$\langle \mathbf{r}_\perp \rangle^\tau = \iint \mathbf{r}_\perp g_z^\tau(\mathbf{r}) dx dy \Big/ \iint g_z^\tau(\mathbf{r}) dx dy, \tag{4}$$

where $\mathbf{r}_\perp = x\hat{\mathbf{x}} + y\hat{\mathbf{y}}$, $\langle \mathbf{r}_\perp \rangle = \langle x \rangle \hat{\mathbf{x}} + \langle y \rangle \hat{\mathbf{y}}$, $g_z(\mathbf{r})$ is the linear momentum density component along z-axis, and superscript $\tau$ = t, o, s refers to total, orbital- and spin-related, respectively. The symmetry of the acoustic vortex field dictates that the centroid shift is definitely equal to zero (i.e., $\langle \mathbf{r}_\perp \rangle^\tau = 0$) when the observation plane is perpendicular to the z-axis.

Here we break this symmetry by tilting the observation plane and further observe the centroid shift of the linear momentum density. When the observation plane is tilted by an angle $\theta$, the coordinate and vector transformation can be realized by a rotation matrix $R(\theta)$. Let the coordinates of the initial and tilted plane be $(x', y', z')$ (i.e., $\mathbf{r}' = x\hat{\mathbf{x}}' + y\hat{\mathbf{y}}' + z\hat{\mathbf{z}}'$) and $(x, y, z)$ (i.e., $\mathbf{r} = x\hat{\mathbf{x}} + y\hat{\mathbf{y}} + z\hat{\mathbf{z}}$), respectively, then one has $\mathbf{r} = R(\theta)\mathbf{r}'$. Therefore, the linear momentum density in the tilted coordinates can be expressed as $\mathbf{g}^\tau(\mathbf{r}) = R(\theta)\mathbf{g}'^\tau(R^{-1}(\theta)\mathbf{r}')$, where $\mathbf{g}'^\tau(\mathbf{r}')$ is the linear momentum density in the initial coordinates. Then the normal linear momentum density in the new coordinates can be obtained as

$$g_z^t = \frac{k_z}{2c_0^2 \rho_0 \omega}\left(k_z \cos\theta + \frac{ly\sin\theta}{\zeta'^2}\right)\left[A_{|l|}(\zeta')\right]^2,$$

$$g_z^s = -\frac{k_z B_{|l|}(\zeta')}{4\rho_0 \omega^3}\left(k_z \cos\theta + \frac{ly\sin\theta}{\zeta'^2}\right), \tag{5}$$

$$g_z^o = \frac{k_z}{2\rho_0 \omega}\left(k_z \cos\theta + \frac{ly\sin\theta}{\zeta'^2}\right)\left(\frac{B_{|l|}(\zeta')}{2\omega^2} + \frac{1}{c_0^2}\left[A_{|l|}(\zeta')\right]^2\right),$$

where $\zeta' = \sqrt{(x\cos\theta - z\sin\theta)^2 + y^2}$. The centroid shift of the normal linear momentum density can be obtained by substituting Eq. (3) into Eq. (5), as follows

$$\langle \mathbf{r}_\perp \rangle^{\mathrm{t}} = \frac{l}{2k_z} \tan\theta \hat{\mathbf{y}},$$

$$\langle \mathbf{r}_\perp \rangle^{\mathrm{s}} = \frac{\alpha l}{2k_z} \tan\theta \hat{\mathbf{y}}, \quad (6)$$

$$\langle \mathbf{r}_\perp \rangle^{\mathrm{o}} = \frac{(1-\alpha)l}{2k_z} \tan\theta \hat{\mathbf{y}},$$

where $\alpha = -c_0^2 \int B_{|l|}(\zeta')\zeta'd\zeta' \Big/ \left\{ 2\omega^2 \int \left[ A_{|l|}(\zeta') \right]^2 \zeta'd\zeta' \right\}$, and $\langle \mathbf{r}_\perp \rangle^{\mathrm{t}} = \langle \mathbf{r}_\perp \rangle^{\mathrm{s}} + \langle \mathbf{r}_\perp \rangle^{\mathrm{o}}$.

Equation (6) is the central result in this Letter. It indicates that all centroid shifts are only along *y*-axis and strictly proportional to *l* and $\tan\theta$ as long as it is observed through ideal vortex beams described by Eq. (1) (note that each step in the derivation process is rigorous and no approximation is involved). Particularly, it is noteworthy that all the centroid shift are inversely proportional to $k_z$ rather than *k*, which provides an important mechanism to amplify this effect by simply adjusting $k_z$, and to enable the experimental observation of this effect in an ultra compact system as will be verified later. Considering that the tilting of the observation plane has no component along *y*-axis, the centroid shifts present to be transverse, which has the flavor of spin and orbital Hall effect. It is noteworthy that different from conventional spin or orbital Hall effect, the transverse shift only results from the tilted observation plane but not actual spin-orbit or orbit-orbit interaction, meaning what we observe is a geometric spin and orbital Hall effect of acoustics.

To further understand its physical origin, we give the explanation from angular momentum perspective. The total, spin and orbital angular momentum density can be expressed as

$$\mathbf{m}(\mathbf{r}) = \mathbf{r} \times \mathbf{g}^{\mathrm{t}}(\mathbf{r}), \qquad \mathbf{s}(\mathbf{r}) = \mathbf{r} \times \mathbf{g}^{\mathrm{s}}(\mathbf{r}), \qquad \mathbf{j}(\mathbf{r}) = \mathbf{r} \times \mathbf{g}^{\mathrm{o}}(\mathbf{r}). \quad (7)$$

The total linear and all three kinds of angular momentum of a cross-section can be obtained by integrating the corresponding momentum density over the whole plane

$$\mathbf{G}^{\mathrm{t}} = \iint \mathbf{g}^{\mathrm{t}}(\mathbf{r})dxdy, \qquad \mathbf{M} = \iint \mathbf{m}(\mathbf{r})dxdy,$$
$$\mathbf{S} = \iint \mathbf{s}(\mathbf{r})dxdy, \qquad \mathbf{J} = \iint \mathbf{j}(\mathbf{r})dxdy. \quad (8)$$

By combining Eqs. (7) and (8), one can obtain

$$M_x = \langle y \rangle^t G_z^t - zG_y^t, \quad M_y = zG_x^t - \langle x \rangle^t G_z^t,$$
$$S_x = \langle y \rangle^s G_z^t - zG_y^s, \quad S_y = zG_x^s - \langle x \rangle^s G_z^t, \qquad (9)$$
$$J_x = \langle y \rangle^o G_z^t - zG_y^o, \quad J_y = zG_x^o - \langle x \rangle^o G_z^t.$$

For simplicity here we set $z = 0$ because we have proved in Eq. (2) that the acoustic linear momentum in $z$-direction is independent of $z$ when $A(\xi, z) = A_{|j|}(\xi)$. Thus Eq. (9) can be further simplified as

$$M_x = \langle y \rangle^t G_z^t, \quad M_y = -\langle x \rangle^t G_z^t,$$
$$S_x = \langle y \rangle^s G_z^t, \quad S_y = -\langle x \rangle^s G_z^t, \qquad (10)$$
$$J_x = \langle y \rangle^o G_z^t, \quad J_y = -\langle x \rangle^o G_z^t.$$

When the observation plane is perpendicular to the propagating direction, three kinds of angular momentum are all in $z$-direction and have no component in other directions (i.e., the transverse angular momentum $M_x = M_y = S_x = S_y = J_x = J_y = 0$), which leads to $\langle x \rangle = \langle y \rangle = 0$. After tilting the observation plane by an angle $\theta$ relative to the $x$-axis, the orientation of the angular momentum is altered, resulting in the emergence of a transverse angular momentum component along the $x$-axis, while the component along the $y$-axis remains invariantly zero. Therefore, in this observation plane, the centroid of linear momentum density shows a shift solely along $y$-axis, with no corresponding shift along $x$-axis. It is also noteworthy in Eq. (10) that the transverse shifts induced by spin and orbital angular momenta are strictly proportional to the transverse component of spin and orbital angular momenta. Furthermore, normal total linear momentum is strictly proportional to the axial wavenumber $k_z$, but the total transverse angular momentum is not related to $k_z$, thus the total transverse shift is inversely proportional to $k_z$ strictly. This allows significantly increasing the transverse shift to amplify geometric spin and orbital Hall effect by adjusting the $k_z$ of the beam, which is beyond attainable with traditional methods for classical waves and substantially simplifies the observation of acoustic giant Hall effect, as will be demonstrated in what follows.

Next, we will numerically calculate the transverse shift for several particular vortex beams satisfying Eq. (1) to verify the above mechanism. Instead of using Gaussian and Laguerre-Gaussian beams with paraxial approximation as in previous works, here our platform for observing acoustic geometric spin and orbital Hall effect is a hard-boundary cylindrical waveguide system that can support diffractionless vortex mode. Noticeably, such a design not only eliminates the influence of interference on centroid shift induced by the diffraction effect but also enables adjusting the radial number $k_\xi$ arbitrarily, which is crucial for the observation of giant acoustic spin and orbital Hall effect in an ultra-compact system by amplifying the transverse shift. By applying the Neumann boundary condition $\partial p/\partial \xi|_{\xi=a} = 0$ to the wave equation $c_0^2 \nabla^2 p(\mathbf{r},t) = \partial^2 p(\mathbf{r},t)/\partial t^2$, a series of discrete modes can be obtained. Beams with one pure mode can be generated by reasonably modulating boundary radius $a$ and sound source, written as $p(\xi,\varphi,z;t) = J_{|l|}(k_\xi \xi)\exp[i(l\varphi + k_z z - \omega t)]$, where $J_l(k_\xi \xi)$ is the $l$th order Bessel function of the first kind [24]. According to Eqs. (2)-(6), the centroid transverse shift of the linear momentum density can be obtained as

$$\langle y \rangle^t = \frac{l}{2k_z}\tan\theta,$$

$$\langle y \rangle^s = \frac{l}{2k_z}\tan\theta \frac{k_\xi^2}{k^2\left(k_\xi^2 a_{l,m}^2 - l^2\right)}, \quad (11)$$

$$\langle y \rangle^o = \frac{l}{2k_z}\tan\theta\left[1 - \frac{k_\xi^2}{k^2\left(k_\xi^2 a_{l,m}^2 - l^2\right)}\right],$$

where $a_{l,m}$ is the $m$th pole of $l$th-order Bessel function of the first kind. The radius of the beam can directly influence the axial wavenumber $k_z$ and further influence the magnitude of linear and angular momenta. Figure 2(a) demonstrates the theoretical and numerical results of three kinds of normalized angular momenta along $x$-axis (the numerator of transverse shift in Eq. (10), e.g., $\langle y \rangle^t = M_x/G_z^t$) and normalized total linear momentum along $z$-axis (the denominator of transverse shift) changing with the boundary radius $R_0$ when the tilting angle $\theta = 45°$ and frequency $f_0 = 2050$ Hz. It shows that the variation of the total linear momentum along $z$-axis with boundary radius is much more rapid than that of angular momenta along $x$-axis, resulting in the decrease of three kinds of transverse shifts $\langle y \rangle^\tau$ with the increase of radius $R_0$, as shown in Fig. 2(b). Figure 2(c) shows the variation curves of three kinds of transverse shifts versus tilting angle $\theta$ when the frequency $f_0 = 2050$ Hz and the radius $R_0 = 5$ cm, while Fig. 2(d) shows the total transverse shift results under different frequencies in response to the tilting angle $\theta$ when the radius $R_0 = 5$ cm. All the numerical results, demonstrated in Fig. 2(b)-(d) which are conducted with a finite element method based on COMSOL MULTIPHYSICS software, are in excellent agreement with our analytical results, demonstrating that the transverse shift can be enhanced by bounding the beam tighter. One can observe from Fig. 2(d) that the total transverse shift of vortex beam can be larger than its radius (5cm) and diameter (10cm) at 2400 Hz and 2100 Hz respectively when the tilting angle reaches 70°, far from achievable with the existing methods.

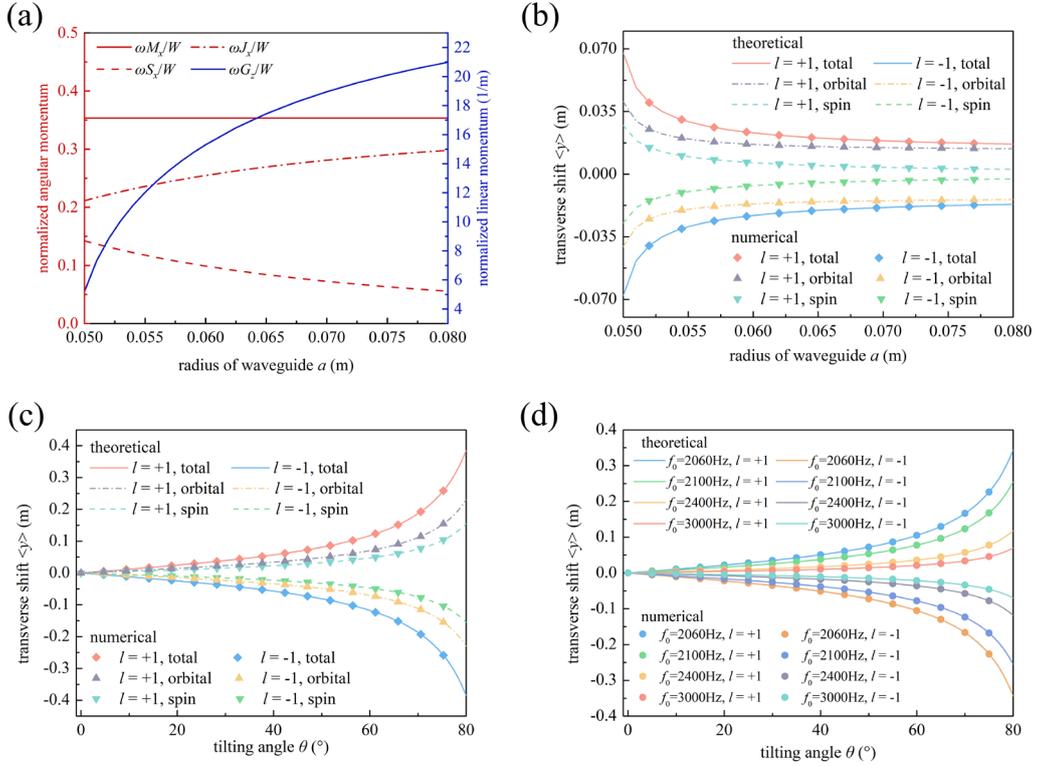

FIG. 2. (a) Numerical results of three kinds of normalized angular momenta along $x$-

axis and normalized total linear momentum along *z*-axis changing as the boundary radius $R_0$ with the tilting angle $\theta = 45°$ and frequency $f_0 = 2050$Hz. Variation of three kinds of transverse shifts when frequency $f_0 = 2050$Hz with (b) the boundary radius $R_0$ with the tilting angle $\theta = 45°$, and (c) the tilting angle $\theta$ when the radius $R_0 = 5$cm. (d) Variation of total transverse shifts under different frequencies with the tilting angle $\theta$ when the radius $R_0 = 5$cm.

Then, we further demonstrate the experimental observation of giant acoustic geometric spin and orbital Hall effect in a cylindrical waveguide made of acrylic resin. In experiment, the radius of the waveguide is 5 cm, the same as in the simulation. Four loudspeakers were placed at one end of the waveguide to generate the vortex field with $l = +1$ and $l = -1$, and sound-absorbing foam was placed at the other end of the waveguide to eliminate the undesired reflection, as shown in Fig. 3(a). We measured the acoustic pressure in a three-dimensional cylindrical region in the middle of the waveguide, based on which we obtained the acoustic pressure at different cross-sections with corresponding tilting angles. Then we calculated the total normal linear momentum density and total transverse shift using Eq. (2) and Eq. (4), respectively. Figure 3(b) shows the theoretical (solid lines) and experimental (solid squares) results of transverse shift of vortex beams at two frequencies of 2100 Hz and 2400 Hz at different cross-sections. Figure 3(c) shows a set of typical results obtained by numerical simulation and experimental measurement. The left and right sub-set are the normal total linear momentum density distribution of the vortex field at frequencies of 2100Hz and 2400Hz in the cross-section with 35° tilting angle, respectively. The experimental results show great agreement with both theoretical and numerical results, demonstrating the significant advantage of our experimental system and observation mechanism. Unlike the system in optics that can only observe the geometric spin Hall effect with small transverse centroid shift yet requires auxiliary polarizer to detect the normal Poynting vector, our system provides a great platform for observing large and pure Hall-effect-induced transverse centroid shift without any auxiliary interfaces interacting with acoustic field, which completely eliminates errors caused by wave-interface interactions and diffraction influence.

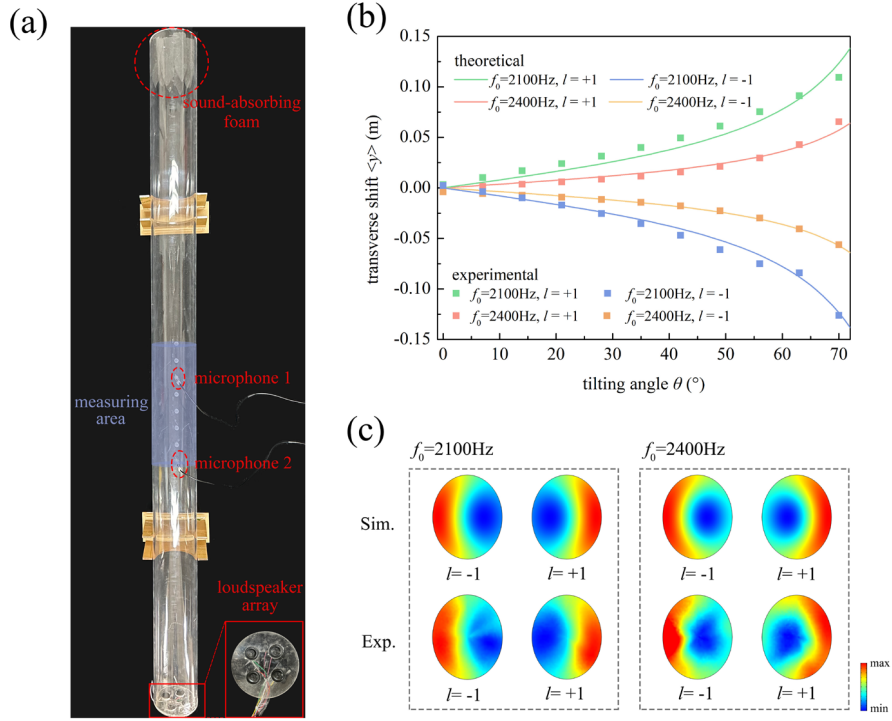

FIG. 3. (a) Photograph of the experimental system. (b) Experimental results of centroid shifts with solid lines being the analytical results predicted by Eq. (11) and the solid squares being the results of experimental measurements. (c) Normal linear momentum density distributions at the cross-section with the tilting angle of 35° from numerical simulation and experimental measurement.

In conclusion, we first theoretically reveal a new acoustic spin-orbit and orbit-orbit effect named geometric spin and orbital Hall effect in which sound-matter interaction is not required, and uncover that this effect can be enhanced by tightly binding beams into small scale by strictly deriving the dependence between transverse shift and $k_z$, which can also eliminate the influence of beams phase/amplitude variation in propagation. Based on these theoretical derivations, we experimentally observe this effect with a cylindrical waveguide whose diameter is smaller than the wavelength of propagating acoustic vortex waves. Extremely large transverse shifts are observed compared with conventional spin/orbital Hall effect. Benefit from our experimental system that does not need any wave-interface interaction, our experimental procedure avoids a large number of systematic errors; our observation method also enlarges the Hall-effect-induced centroid shift and eliminates the extra interferential centroid shifts that are induced by other factors at the same time. The combination of these two advantages allows us to observe the spin and orbital Hall effect in longitudinal acoustic system that lacks precise observational methods such as quantum weak measurements, and the theoretical and experimental results agree very well, even far better than those in the observation of geometric spin Hall effect in optics. Our work provides deeper insights into the physics underlying the spin and orbital angular momentum of acoustic vortex beams and may help the development of vortex-based novel acoustic devices.

**Acknowledgements**

This work was supported by National Key R&D Program of China (Grant Nos. 2022YFA1404402 and 2017YFA0303700), the National Natural Science Foundation of China (Grant Nos. 12304493, 11634006 and 12174190), the Natural Science Foundation of Jiangsu Province (Grant No. BK 20230767), a project funded by the Priority Academic Program Development of Jiangsu Higher Education Institutions and High-Performance Computing Center of Collaborative Innovation Center of Advanced Microstructures.